\documentstyle[11pt,aasms4]{article} 
\input{psfig.tex} 
\lefthead{Barrow \& Magueijo} 
\righthead{Can a changing $\alpha$ explain ... } 
 
\begin{document} 
 
\title{Can a changing $\alpha$ explain the Supernovae results?} 
\author{John D. Barrow$^1$ and Jo\~ao Magueijo$^2$} 
\affil{\  
$^1$DAMTP, Centre for Mathematical Sciences, Cambridge University, 
Wilberforce Rd., Cambridge CB3 0LA, U.K.\\$^2$%
Theoretical Physics Group, The Blackett Laboratory, Imperial College, 
Prince Consort Rd., London, SW7 2BZ, U.K.}

\begin{abstract} 
We show that the Supernovae results, implying evidence for an accelerating 
Universe, may be closely related to the recent discovery of redshift 
dependence in the fine structure constant $\alpha $. The link is a class of 
varying speed of light (VSL) theories which contain cosmological solutions 
similar to quintessence. During the radiation dominated epoch the 
cosmological constant $\Lambda $ is prevented from dominating the Universe 
by the usual VSL mechanism. In the matter epoch the varying $c$ effects 
switch off, allowing $\Lambda $ to eventually surface and lead to an 
accelerating Universe. By the time this happens the residual variations in $c 
$ imply a changing $\alpha $ at a rate that is in agreement with 
observations. 
\end{abstract} 


\tighten

 
\section{Introduction} 
 
Two puzzling observations are challenging cosmologists. The Supernovae 
Cosmology Project and the High-z Supernova Search
(\cite{super,super1,super2,super3}) have
extended the reach of the Hubble diagram to high redshift and provided new 
evidence that the expansion of the universe is accelerating. 
This {\it may} imply 
that there exists a significant positive cosmological constant, $\Lambda $. 
In separate work, the spacings between quasar (QSO) 
absorption lines were examined 
in Keck I data at medium redshifts, $z\sim 1$, (\cite{alpha}) and compared 
with those in the laboratory (see also \cite{alpha1,alpha2,alpha3,alpha4}). 
These observations are sensitive to time variations in the value of the fine 
structure constant $\alpha =e^2/(\hbar c)$, (where $e$ is the electron 
charge, $\hbar $ Planck's constant, and $c$ the speed of light), at a rate 
one million times slower that the expansion rate of the universe. 
Evidence was found  for a small 
variation in the value of $\alpha $ at redshifts $\sim 1$. This could be 
produced by intrinsic time variation or by some unidentified line-blending 
effect. In this Letter we assume that the variation is intrinsic and show 
that there may be a link between the observations of cosmological 
acceleration and varying $\alpha $. 
 
If $\Lambda >0$, then cosmology faces a very serious fine tuning problem, 
and this has motivated extensive theoretical work. There is no 
theoretical motivation for a value of $\Lambda $ of currently
 observable magnitude; a 
value $10^{120}$ times smaller that the 'natural' Planck scale of density is 
needed if $\Lambda $ becomes important near the present time. Such a small 
non-zero value of $\Lambda $ is 'unnatural' in the sense that making it zero  
{\it reduces} the symmetry of spacetime. A tentative solution is quintessence 
(\cite{quint}): the idea that $\Lambda $ might be a rolling scalar field 
exhibiting very long transients. Here we introduce another explanation. 
 
There are a variety of possible physical expressions of a changing $\alpha $%
. Bekenstein proposed a varying $e$ theory (\cite{bek}). An alternative is 
the varying speed of light (VSL) theory (\cite{mof,vsl0,vsl1}) in which 
varying $\alpha $ is expressed as a variation of the speed of light. The 
choice between these two types of theory transcends experiment, and merely 
reflects theoretical convenience in the choice of units (\cite{vsl2}). The 
simplest cosmology following from VSL is known to contain an attractor in 
which $\Lambda $ and matter remain at fixed density ratios throughout the 
life of the universe (\cite{vsl3}). Such attractor solves the fine tuning 
problem forced upon us by the supernovae results. Hence there is scope for 
the observed changing $\alpha $ to be related to the observed acceleration 
of the universe. In this {\it Letter} we propose a model which leads to 
good quantitative agreement, given experimental errors, between the 
observations of acceleration and varying $\alpha $. In Section~\ref{hubble} 
we examine the construction of the Hubble diagram in VSL theories, and the 
interpretation of varying-$\alpha $ experiments. Then in Section~\ref{model} 
we present an example of a VSL model which can jointly explain the 
supernovae results and the Webb et al varying-$\alpha $ results. We conclude 
with a discussion of some further aspects of the model proposed, to be 
investigated elsewhere. 
 
\section{The VSL Hubble diagram} 
 
\label{hubble} The Hubble diagram is a plot of luminosity distance against 
redshift. The purpose is to map the expansion factor $a(t)$, where $t$ is 
the comoving proper time. Redshifts provide a measurement of $a$ at the time 
of emission. If the objects under observation are ``standard candles'' (as 
Type Ia supernovae are assumed to be), their apparent brightness gives their 
(luminosity) distance, which, if we know $c$, tells us their age. By looking 
at progressively more distant objects we can therefore map the curve $a(t)$. 
 
We now examine how this construction is affected by a changing $c$. In  
\cite{vsl0} we showed that $E\propto c^2$ for photons in free flight.  
We also showed that quantum mechanics remains unaffected by a changing 
$c$ if $\hbar\propto c$ (in the sense that quantum numbers are adiabatic 
invariants). Then all relativistic energies scale like $c^2$.
If for non-relativistic systems $\hbar \propto 1/c$, the Rydberg 
energy $E_R=m_e e^4/(2\hbar ^2)$ also scales like $c^2$.
Hence all absorption lines, ignoring the fine structure, scale like $c^2$. 
When we compare lines from near and far systems we should therefore see no 
effects due to a varying $c$; the redshift $z$ is still  
$1+z_e=a_o/a_e  $,
where $o$ and $e$ label epochs of observation and emission.
 
In order to examine luminosity distances, we need to reassess the concept of 
standard candles. For simplicity let us first treat them as black bodies. 
Then their temperature 
scales as $T\propto c^2$ (\cite{vsl0}), their energy density scales as $\rho 
\propto T^4/(\hbar c)^3\propto c^2$, and their emission power as $P=\rho 
/c\propto c$, implying that standard candles are brighter in the early 
universe if $\dot c<0$. However, the power emitted by these candles, in free 
flight, scales like $c$; each photon's energy scales like $c^2$, its speed 
like $c$, and therefore its energy flux like $c$. The received flux, as 
a function of $c$, therefore scales like: 
\begin{equation} 
P_r={\frac{P_ec^2}{4\pi r^2c}}\propto c 
\end{equation} 
where $r$ is the conformal distance to the emitting object, and the  
subscripts $r$ and $e$ label received and emitted.  
In an expanding universe we therefore still have  
\begin{equation} 
P_r={\frac{P_{e0}}{4\pi r^2a_0^2}}{\left( \frac a{a_o}\right) }^2, 
\label{lum} 
\end{equation} 
where $P_{e0}$ is the emitting power of standard candles today. 
Notice that the above argument is still valid if the candles 
are not black bodies; it depends only on the scaling properties 
of emitted and received power. 
 
We can now set up the Hubble diagram. Consider the Taylor expansion  
\begin{equation} 
a(t)=a_0[1+H_0(t-t_0)-{\frac 12}q_0H_0^2(t-t_0)^2+...]  \label{expansion} 
\end{equation} 
where $H_0=\dot a_0/ a_0$ is the Hubble constant, and 
$q_0=-\ddot a_0 a_0/ \dot a_0^2$ is the decceleration parameter. 
Hence up to second order 
$z=H_0(t_0-t)+(1+q_0/2)H_0^2(t-t_0)^2$, 
or  
\begin{equation} 
t_0-t={\frac 1{H_0}}[z-(1+q_0/2)z^2+...].  \label{t-t0} 
\end{equation} 
From (\ref{lum}) we find that the luminosity distance $d_L$ is  
\begin{equation} 
d_L={\left( \frac{P_{e0}}{4\pi P_0}\right) }^{1/2}=a_0^2{\frac ra}%
=a_0r(1+z_e).
\end{equation} 
The conformal distance to the emitting object is given by  
$r=\int_t^{t_0}{c({t})d{t}}/{a({t})}.$ 
From (\ref{expansion}) we have that  
\begin{equation} 
r=c_0[(t_0-t)+{\frac{1-n}2}H_0(t_0-t)^2+...] 
\end{equation} 
where we have assumed that locally $c=c_0a^n$ (that is  
$c=c_0[1+nH_0(t-t_0)+...]$). Substituting (\ref{t-t0}) we 
finally have \footnote{Had we assumed that $\hbar \propto c$ for
all systems we would have got instead 
$d_L=(c_0/{\tilde H}_0)[z+{\frac 12}(1-(q_0(1+4n)+n))z^2]$,
with ${\tilde H}_0=(1-4n)H_0$. This does not affect any of the conclusions.
}:  
\begin{equation} 
d_L={\frac{c_0}{H_0}}[z+{\frac 12}(1-(q_0+n))z^2+...] 
\end{equation} 
We see that besides the direct effects of VSL upon the expansion rate of the 
universe, it also induces an effective{\it \ acceleration} in the Hubble 
diagram as an ``optical illusion'' (we are assuming that $c$ decreases in 
time: $n<0$). This is easy to understand. We have seen that VSL introduces 
no intrinsic effects in the redshifting spectral line 
or in the dimming of standard 
candles with distance and expansion.  
The only effect VSL induces on the 
construction of the Hubble diagram is that for the same redshift (that is, 
the same distance into the past) objects are farther away from us because 
light travelled faster in the past. But an excess luminosity distance, for 
the same redshift, is precisely the hallmark of cosmological acceleration. 
However, we need to consider the other experimental input to our work: the  
\cite{alpha} results. By measuring the fine structure in absorption systems 
at redshifts $z\sim O(1)$ we can also map the curve $c(t)$. Since $%
c=c_0[1+nH_0(t-t_0)+...]$ we have $c=c_0[1-nz+...]$, and so to first order $%
\alpha =\alpha _0[1+2nz+...]$. However, the results presented in  \cite 
{alpha} show that $n$ is at most of order $10^{-5}.$ This means that the 
direct effects of varying $c$ permitted by the QSO absorption system 
observations are far too small to explain the observed acceleration. We need 
to look at a fully self-consistent generalisation of general relativity 
containing the scope for varying $c$.
 
\section{The model} 
\label{model} We start with some general properties of the dynamics of $c$.
Drawing inspiration from dilaton 
theories (like Brans-Dicke gravity) we take  
$\psi =\log (c/c_0) $ 
as the dynamical field associated with $c$. 
Indeed, powers of $c$ appear in all coupling constants, which in turn can be 
written as $e^\phi $, where $\phi $ is the dilaton. Another theory using
a similar dynamical variable is the changing $\alpha$ theory of \cite{bek}
(which uses $\log(\alpha)$).

We then endow $\psi $ with a dynamics similar to the dilaton.
The left-hand side for the $\psi $ 
equation should therefore be $\Box\psi$ (in the preferred Lorentz
frame - to be
identified with the cosmological frame). This structure ensures that the 
propagation equation for $\psi $ is second-order and hyperbolic, i.e. 
propagation is causal. Since VSL  breaks 
Lorentz invariance other expressions
would be possible - but then the field $\psi$ would propagate non-causally.
An example is $(g_{\mu\nu}+u_\mu u_\nu)(\nabla_\mu\psi)(\nabla_\nu\psi)$,
where $u^\mu$ is the tangent vector
of the local preferred frame.

On the other hand one need not choose (as in Brans-Dicke theories) 
the source term to be $\rho -3p$, where $\rho $ and $p$ are the energy 
density and pressure of matter respectively. Without the requirement
of Lorentz invariance other expressions are possible, and using them 
does not conflict with local causality. If  $T^{\mu\nu}$ is the 
stress-energy tensor we can choose as a source term
$T^{\mu\nu}(g_{\mu\nu}+u_\mu u_\nu)$;
 that is changes in $c$ are driven by the matter
pressure. We find this choice is the one that gives interesting effects.

For a homogeneous field in an expanding universe we therefore have
${\ddot \psi }+3(\dot a/a){\dot \psi }=4\pi G\omega p/{c^2}$, 
where $p$ is the total pressure of the matter fields and $\omega $ is a 
coupling constant (distinct from the Brans Dicke coupling constant).  
The full self-consistent system of equations in a 
matter-plus-radiation universe containing a cosmological constant stress 
%
$\rho _{\Lambda }=(\Lambda c^2)/(8\pi G)$   
is therefore  
\begin{eqnarray} 
\ddot \psi +3{\frac{\dot a}a}\dot \psi &=&4\pi G\omega {\frac{\rho _\gamma }{%
3\label{totaleqns1}},} \label{sor}\\ 
\dot \rho _\gamma +4{\frac{\dot a}a}\rho _\gamma &=&-2\rho _\Lambda \dot 
\psi ,  \label{totaleqns2} \\ 
\dot \rho _\Lambda &=&2\rho _\Lambda \dot \psi ,  \label{totaleqns3} \\ 
\dot \rho _m+3{\frac{\dot a}a}\rho _m &=&0,  \label{totaleqns4} \\ 
{\left( \frac{\dot a}a\right) }^2 &=&{\frac{8\pi G}3}(\rho _m+\rho _\gamma 
+\rho _\Lambda ),  \label{totaleqns5} 
\end{eqnarray} 
where subscripts $\gamma $ and $m$ denote radiation and matter respectively. 
We have assumed that the sink term (\ref{totaleqns3}) is reflected in 
a source term in (\ref{totaleqns2}) (and not in (\ref{totaleqns4})). 
This is due to the fact that this term is only significant very early on, 
when even massive particles behave like radiation.  
We have ignored curvature terms because in the quasi-lambda dominated 
solutions we are about to explore we know that these are smaller than $%
\rho _{\Lambda  
}$, (\cite{vsl3}).   
Here, in complete contrast to 
Brans-Dicke theory, the field $\psi $ is only driven by radiation pressure 
in the dust-dominated era. In other words, only conformally invariant 
forms of matter couple to the field $\psi$.

\begin{figure}
\centerline{{\psfig{file=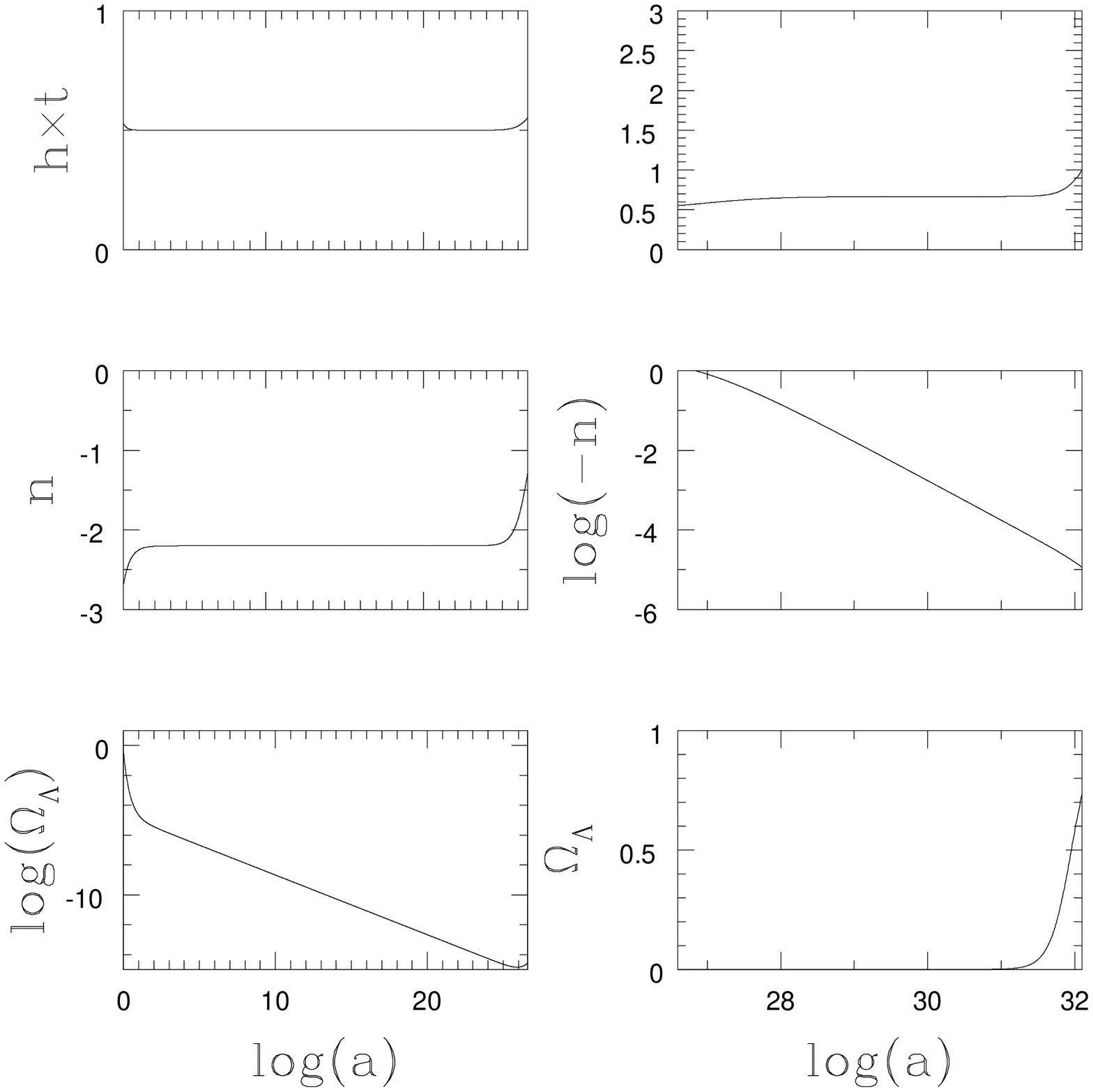,width=12 cm,angle=0}}} 
\caption{Evolution of $t\times h$ (where $h= {\frac{\dot a}{a}}$), 
$n={\frac{\dot c/c}{\dot 
a /a}}$, and $\Omega_\Lambda$ in log or linear plots as appropriate. The 
panels on the left (right) describe the radiation (matter) dominated epoch. 
We have taken $\omega=-4.4$ (a modest $10\%$ tuning over the critical value
$\omega=-4$). In the radiation epoch $n=-2.2$, 
$\Omega_\Lambda$ slowly drops to zero, and 
the expansion factor has the usual dependence $a\propto t^{1/2}$.  
As the Universe enters the matter epoch $n$ starts dropping towards zero, 
$a\propto t^{2/3}$, and then  
$\Omega_\Lambda$ starts to increase. Eventually $n$ is of order $10^{-5}$  
and $\Omega_\Lambda\approx 0.7$. 
This type of behaviour occurs for a large, non finely tuned, 
region of couplings $\omega$ and initial conditions (here 
$\Omega_\Lambda=\Omega_\gamma=0.5$). } 
\label{fig1} 
\end{figure}

In a radiation-dominated universe the behaviour of this system changes at 
the critical value $\omega =-4$. For $\omega <-4$ we reach a flat 
$\rho_\Lambda 
=0 $ attractor as $t\rightarrow \infty $. For $-4<\omega <0$ we have 
attractors for which $\rho _\Lambda $ and $\rho _\gamma $ maintain a 
constant ratio (see \cite{vsl3}). In Fig.~\ref{fig1} we plot a numerical 
solution to 
this system, with $\omega =-4.4$ (a $10\%$ tuning below the critical value $%
\omega =-4$) and $n=-2.2$ during the radiation epoch. As expected from \cite 
{vsl3}, this forces $\Omega _\Lambda $ to drop to zero, while the expansion 
factor acquires a radiation-dominated form, with $a\propto t^{1/2}$. By the 
time the matter-dominated epoch is reached, $\Omega _\Lambda $ is of order $%
10^{-12}$. During the matter epoch, the source term for $\psi $ disappears 
in eq. (\ref{sor}), $n$ starts to approach zero, $\Omega _\Lambda $ starts 
to increase, and the expansion factor takes on the $a\propto t^{2/3}$ 
dependence of a matter-dominated universe. A few expansion times into the 
matter epoch, $\Omega _\Lambda $ becomes of order 1 and the universe begins 
accelerating. By the time this happens $n$ is of order $10^{-5}$, in 
agreement with the expectations of \cite{alpha}. This type of behaviour 
can be achieved generically, for different initial conditions, with a tuning 
of $\omega $ that never needs to be finer than a few percent. 
 
We can provide an approximate argument explaining why this theory should 
display this type of behaviour and why we need so little fine tuning of $%
\omega $ to explain the supernovae experiments. If we neglect changes in $c$ 
after matter-radiation equality, $t_{eq}$, we are going to require  
${\rho _\Lambda (t_{eq})}/{\rho (t_{eq})}\approx z_{eq}^{-3}\sim 
10^{-12}$. 
Let $c=c_0a^{n(t)}$, with $n=-2-\delta $, and $n=\omega /2$ during the 
radiation epoch. We can integrate the conservation equations to give  
\begin{equation} 
{\frac \rho {\rho _\Lambda }}={\frac A{a^4\rho _\Lambda }}-{\frac n{n+2},} 
\end{equation} 
with $A$ constant, from which it follows that  
\begin{equation} 
{\frac \rho {\rho _\Lambda }}={\frac 2\delta }{\left[ {\left( 1+{\frac 
\delta 2}{\frac{\rho _i}{\rho _{\Lambda i}}}\right) }{\left( \frac 
a{a_i}\right) }^{2\delta }-1\right] .} 
\end{equation} 
We see that assymptotically $\rho/\rho_\Lambda$ grows to infinity, 
if $\delta >0$ (the flat $\rho_\Lambda=0$ attractor of \cite{vsl3}). 
However the growth is very slow even if $\delta$ is not very small. 
Our theory displays very long transients, and a very slow convergence 
to its attractor, a property similar to quintessence models 
(\cite{quint}).  
It is therefore possible to achieve $\rho _\Lambda /\rho \sim 10^{-12}$ at 
the end of the radiation epoch, with $\delta $ chosen to be of order $0.1$.

Now, why is the change in $c$ of the right order of magnitude to explain the 
results of \cite{alpha}? With a solution of the form $c=c_0a^{n(t)}$ we 
find that  
\begin{equation} 
n(t)\approx {\frac{\omega \rho _\gamma }{3(\rho _m+2\rho _\Lambda )}} 
\end{equation} 
valid in the matter dominated era, regardless of the details of the
radiation to matter transition. 
With $\omega \approx -4$ we therefore have  
\begin{equation} 
n(t_0)\approx -{\frac 43}{\frac{2.3\times 10^{-5}}{h^2(1+\Omega _\Lambda )}} 
\end{equation} 
of the right order of magnitude. The order of magnitude of the index $n\sim 
10^{-5}$, observed by \cite{alpha}, is therefore fixed by the ratio of the 
radiation and the matter energy densities today. 
 
\section{Discussion} 
 
In this Letter we proposed a theory relating the supernovae results and the 
observations by \cite{alpha}.
The theory we have proposed is one example within a 
class whose members exhibit similar behaviour. In these theories the 
gravitational effect of the 
pressure drives changes in $c$, and these convert the energy density in $%
\Lambda $ into radiation. Thus 
$\rho_\Lambda $ is prevented from dominating the 
universe during the radiation epoch. As the universe cools down, massive 
particles eventually become the source of pressureless matter and create a 
matter-dominated epoch. In the matter epoch the variation in $c$ comes to a 
halt, with residual effects at $z\approx 1-5$ at the level observed by Webb 
et al. As the $c$ variation is switched off, the $\Lambda $ stress 
resurfaces, and dominates the universe for a few expansion times in the 
matter-dominated era, in agreement with the supernovae results.

In a forthcoming publication we shall address other aspects of this
theory, beyond the scope of this Letter. We mention nucleosynthesis,
the location in time of a quantum epoch, and perturbations around the
homogeneous solution discussed here (see \cite{ot}). Nucleosynthesis
in particular may provide significant constraints on this class of models.
However we expect a variation in $\alpha$ to require variations in
other couplings if some unification exists.  Nucleosynthesis
involves many competing effects contributed by weak, strong and
electromagnetic, and gravitational interactions and we do not know how to
incorporate all the effects self-consistently. 
Studies of the effects of varying constants coupled by Kaluza-Klein extra
dimensions have been made by \cite{kkk} and \cite{alpha4}.
The most detailed study to date was conducted by \cite{nuc}.

\section*{Acknowledgements} 
 
JDB is partially 
supported by a PPARC Senior Fellowship. 
JM would like to thank K. Baskerville and D. Sington for help with  
this project.

\newpage  

\end{document}